\documentclass{ws-mpla}

\begin{document}
\newcommand{\om }{\omega}
\newcommand{\lam}{\lambda}
\newcommand{\gam}{\gamma}
\markboth{Jen-Chieh Peng}
{Neutron Electric Dipole Moment Experiments}

\catchline{}{}{}{}{}

\title{Neutron Electric Dipole Moment Experiments
}

\author{Jen-Chieh Peng}

\address{Department of Physics, University of Illinois, 1110 W. Green St., Urbana, IL,
61801 U.S.A.\\
jcpeng@uiuc.edu}

\maketitle

\pub{Received (Day Month Year)}{Revised (Day Month Year)}

\begin{abstract}

The neutron electric dipole moment (EDM) provides unique information on
CP violation and physics beyond the Standard Model. We first review 
the history of experimental searches for neutron
electric dipole moment. The status of future neutron EDM experiments,
including experiments using ultra-cold neutrons produced in
superfluid helium, will then be presented.

\end{abstract}

\ccode{PACS Nos.: 21.10.Ky, 11.30.Er, 13.40.Em}

\section{Introduction}	

The possibility of a non-zero value for neutron EDM continues to be
of central importance in physics and cosmology. A non-zero neutron
EDM would be a direct evidence for time-reversal symmetry violation. It also
implies CP-violation under the assumption of CPT invariance.
To date, only two examples of CP-violation have been found: decays of
neutral K mesons and B mesons. CP-violation is believed to
have occured during the Big Bang baryogenesis that led to the present
matter-antimatter asymmetry in the Universe. Although CP-violation
observed in K and B meson decays can be incorporated phenomenologically
within the Standard Model, the strength of
the effect is not large enough to explain the matter-antimatter asymmetry.
It is likely that a full description of CP-violation would invoke
non-standard models, many of which predict a neutron EDM large enough to be
accessed experimentally\cite{lamoreaux}. 
Therefore, a sensitive measurement of the neutron
EDM is of fundamental interest since it could identify new sources of
CP-violation as well as physics beyond the Standard Model.

In this article, we first review the history of neutron EDM experiments.
The prospect for future neutron EDM experiments will then be presented. In
particular, the status of proposed experiments using ultra cold neutrons at
various laboratories will be discussed.

\section{Existing Neutron EDM Experiments}

The history of neutron EDM measurements is closely connected to the
development of our knowledge on discrete symmetries in physics.
In 1950, when parity was considered as an inviolable
symmetry, Purcell and Ramsey\cite{purcell50} pointed out
the need to test this symmetry via a
detection of neutron EDM. They then carried out a pioneering
experiment\cite{smith51,smith57}
setting an upper limit at $5 \times 10^{-20} e \cdot cm$ for neutron EDM.
The role of baryon (proton, neutron,
hyperons) EDM in testing parity symmetry was extensively discussed in
the seminal paper of Lee and Yang\cite{lee56}, who cited
the yet-unpublished neutron
EDM result from Smith, Purcell, and Ramsey\cite{smith51,ramsey56}.

The discovery of parity violation in 1957\cite{wu57}
prompted Smith et al. to publish
their neutron EDM result\cite{smith57}. By this time,
however, it was recognized\cite{landau57,zeldo57} that
time-reversal invariance would also prevent the neutron from possessing
a non-zero EDM. Since no evidence of T-violation was found even in systems
which exhibited maximal parity violation, a non-zero neutron EDM was
regarded as highly unlikely. The experimental activities
on neutron EDM therefore
lay dormant until CP-violation, directly linked to T-violation via the CPT
theorem\cite{luders54},
was discovered in 1964\cite{christ64}.

The interest in neutron EDM was greatly revived when a large number
of theoretical models, designed to account for the CP-violation phenomenon
in neutral kaon decays, predicted a neutron EDM large enough to be detected.
Many ingenious technical innovations have since been implemented, and
the experimental limit of neutron EDM was pushed down to $2.9 \times 
10^{-26} e\cdot cm$,
a six order-of-magnitude improvement over the first EDM experiment.
Unlike parity-violation, the underlying physics for CP-violation
remains a great enigma nearly 45 years after its discovery. Improved
neutron EDM measurements will continue to provide
stringent tests for various theoretical models and to help reveal the
origin of CP-violation.

\begin{figure}[th]
\centerline{\psfig{file=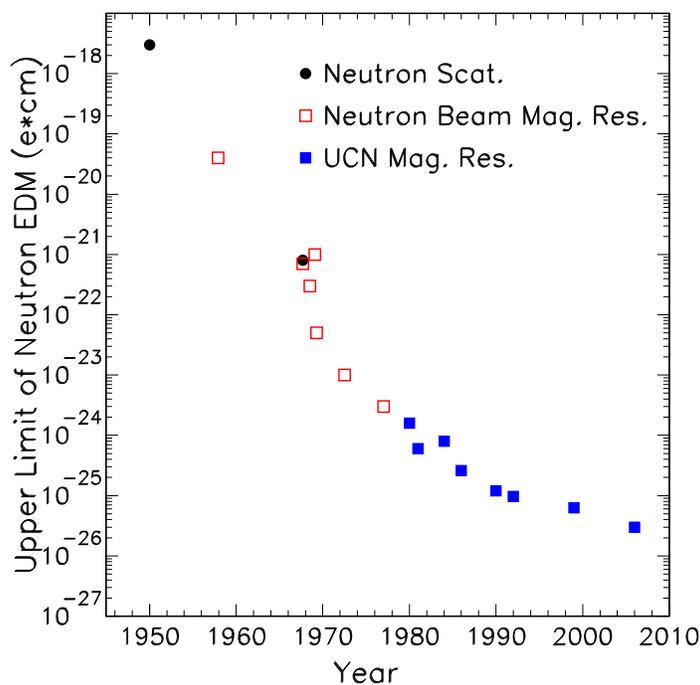,width=4.0in}}
\caption{Upper limits of neutron EDM plotted as a function of year
of publication.}
\label{fig:edmhis1}
\end{figure}

Table~\ref{tab:edmhis1} lists the results
from existing neutron EDM experiments. In Figure~\ref{fig:edmhis1}
the neutron EDM upper limits are plotted versus year of publication.
The different symbols
in Figure~\ref{fig:edmhis1} signify different experimental techniques,
which fall into three
categories. The first one, which consists of only two experiments,
utilizes neutron scattering to probe the effect of neutron EDM.
The second and
third categories both involve magnetic resonance technique. In the presence of
a strong external electric field, a finite neutron EDM would cause
a shift of the magnetic resonance frequency. From 1950 to mid 1970's,
thermal or cold neutron beams have been used for the
measurements (category II). Since early 1980's, all neutron EDM experiments
have utilized ultra-cold neutrons (UCNs), which provide
the most sensitive measurements to date (category III).

\subsection{Neutron EDM from neutron scattering}

The upper limit of the neutron EDM was first determined in 1950 by
Purcell and Ramsey\cite{purcell50} from an analysis of earlier
experiments of neutron-nucleus scattering\cite{havens47,fermi47}.
In these experiments, the strength of neutron-electron interaction
was deduced from the interference between the neutron-nucleus
and neutron-electron scattering. If the observed neutron-electron
interaction strength is attributed entirely to the neutron EDM ($d_n$), an
upper limit of $d_n \le 3 \times 10^{-18} e \cdot cm$ is obtained.

Another technique to search for the neutron EDM
is the Bragg reflection of thermal neutrons from a single
crystal. The scattering amplitude of thermal neutrons comes mainly from
the nuclear interaction. However, the Coulomb field exerted by
the positively charged nucleus on the incident neutron can
provide additional contributions. First, it produces an effective
magnetic field of $\vec v \times \vec E$ in the neutron rest frame.
The neutron magnetic moment interacts with this magnetic
field and leads to the Schwinger scattering. The effect of Schwinger
scattering is maximal when the neutron polarization is perpendicular to
the scattering plane. If the neutron has a non-zero EDM, the Coulomb field
of the nucleus would lead to an additional potential
$V_d(r) = - \vec d_n \cdot \vec E(r)$, where $\vec d_n$ is
the neutron EDM. The effect of the neutron EDM is maximal
when the neutron polarization lies on the scattering plane. This
feature allows the isolation of the EDM effect from
the Schwinger scattering.

\vfill
\eject

\begin{table}[tbh]
\tbl{Summary of neutron EDM experiments.}
{\begin{tabular}{|c|c|c|c|c|}
\hline
Exp. Type & $\langle v \rangle$ & E (KV/cm) &
Coh. time & EDM  \\
(Lab, year) & (m/sec) & B (Gauss) &
(second) & ($e \cdot cm$) \\
\hline\hline
Scattering & 2200 & $\sim 10^{15}$ & $\sim 10^{-20}$ & \\
(ANL, 1950)\cite{purcell50,fermi47} & & - & & $< 3 \times 10^{-18}$ \\
\hline
Beam Mag. Res. & 2050 & 71.6 & 0.00077 & $(-0.1 \pm 2.4)$
$\times 10^{-20}$ \\
(ORNL, 1957)\cite{smith57} & & 150 & & $< 4 \times 10^{-20}$ (90\% C.L.) \\
\hline
Beam Mag. Res. & 60 & 140 & 0.014 & $(-2 \pm 3)$ $\times 10^{-22}$ \\
(ORNL, 1967)\cite{miller67} & & 9 & &$ <7 \times 10^{-22}$ (90\% C.L.) \\
\hline
Bragg Reflection & 2200 & $\sim 10^9$ & $\sim 10^{-7}$ &
$(2.4 \pm 3.9)$ $\times 10^{-22}$ \\
(MIT, 1967)\cite{shull67} & & - & & $ <8 \times 10^{-22}$ (90\% C.L.) \\
\hline
Beam Mag. Res. & 130 & 140 & 0.00625 & $(-0.3 \pm 0.8)$ $\times 10^{-22}$ \\
(ORNL, 1968)\cite{dress68} & & 9 & &$< 3 \times 10^{-22}$ \\
\hline
Beam Mag. Res. & 2200 & 50 & 0.0009 & \\
(BNL, 1969)\cite{cohen69} & & 1.5 & &$< 1 \times 10^{-21}$ \\
\hline
Beam Mag. Res. & 115 & 120 & 0.015 & $(1.54 \pm 1.12) \times 10^{-23}$ \\
(ORNL, 1969)\cite{baird69} & & 17 & &$< 5 \times 10^{-23}$ \\
\hline
Beam Mag. Res. & 154 & 120 & 0.012 & $(3.2 \pm 7.5) \times 10^{-24}$ \\
(ORNL, 1973)\cite{dress73} & & 14 & &$< 1 \times 10^{-23}$ (80\% C.L.) \\
\hline
Beam Mag. Res. & 154 & 100 & 0.0125 & $(0.4 \pm 1.5) \times 10^{-24}$ \\
(ILL, 1977)\cite{dress77} & & 17 & &$< 3 \times 10^{-24}$ (90\% C.L.) \\
\hline
UCN Mag. Res.& $\le$ 6.9 & 25 & 5  & $(0.4 \pm 0.75) \times 10^{-24}$ \\
(PNPI, 1980)\cite{altarev80} & & 0.028 & & $< 1.6 \times 10^{-24}$ (90\% C.L.) \\
\hline
UCN Mag. Res. & $\le$ 6.9 & 20 & 5 & $(2.1 \pm 2.4) \times 10^{-25}$ \\
(PNPI, 1981)\cite{altarev81} & & 0.025 & &$< 6 \times 10^{-25}$ (90\% C.L.) \\
\hline
UCN Mag. Res. & $\le$ 6.9 & 10 & 60-80 & $(0.3 \pm 4.8)
\times 10^{-25}$ \\
(ILL, 1984)\cite{pendlebury84} & & 0.01 & &$< 8 \times 10^{-25}$ (90\% C.L.) \\
\hline
UCN Mag. Res. & $\le$ 6.9 & 12-15 & 50-55 & $-(1.4 \pm 0.6)
\times 10^{-25}$ \\
(PNPI, 1986)\cite{altarev86} & & 0.025 & &$< 2.6 \times 10^{-25}$ (95\% C.L.) \\
\hline
UCN Mag. Res. & $\le$ 6.9 & 16 & 70 & $-(3 \pm 5) \times 10^{-26}$ \\
(ILL, 1990)\cite{smith90} & & 0.01 & &$< 12 \times 10^{-26}$ (95\% C.L.) \\
\hline
UCN Mag. Res. & $\le$ 6.9 & 12-15 & 70-100 & $(2.6 \pm 4.5)
\times 10^{-26}$ \\
(PNPI, 1992)\cite{altarev92} & & 0.018 & & $< 9.7 \times 10^{-26}$ (90\% C.L.) \\
\hline
UCN Mag. Res. & $\le$ 6.9 & 4.5 & 120-150 & $(-1 \pm 3.6)
\times 10^{-26}$ \\
(ILL, 1999)\cite{harris99} & & 0.01 & &$< 6.3 \times 10^{-26}$ (90\% C.L.) \\
\hline
\end{tabular}}
\label{tab:edmhis1}
\end{table}

Shull and Nathan\cite{shull67} measured 
Bragg reflection of polarized neutrons off a CdS crystal, and
they obtained an upper limit for the neutron EDM as
$5 \times 10^{-22} e \cdot cm$. An important limitation of the crystal 
reflection method is the difficulty
to align the crystal orientation along the
polarization direction of the incident neutrons. Any residual misalignment
would allow the Schwinger scattering to contribute 
in a fashion similar to neutron EDM. The limit on $d_n$ of the 
Shull and Nathan experiment is consistent with an misalignment angle of
$1.6 \pm 1.0$ milliradian.

\subsection{Neutron EDM from in-flight neutron magnetic resonance}

The method used in this type of measurements is the magnetic
resonance technique invented by Alvarez and Bloch\cite{alvarez40}. 
For transversely
polarized neutrons traversing a region of uniform magnetic field
$\vec B_0$ and an electric field $\vec E_0$ parallel to $\vec B_0$,
the precession frequency ($\nu$) is given by
\begin{equation}
h \nu = -2 \mu B_0 -2 d_n E_0,
\label{eq:larmor1}
\end{equation}
where $\mu$ is the neutron magnetic moment and $d_n$ the neutron
EDM. Upon reversal of the electric field direction, the
precession frequency will shift by
\begin{equation}
h \Delta \nu = -4 d_n E_0.
\label{eq:larmor2}
\end{equation}
Therefore, the neutron EDM can be determined as
\begin{equation}
d_n = - \frac {h \Delta \nu} {4 E_0}.
\label{eq:larmor3}
\end{equation}
The neutron precession frequency can be accurately measured using the
technique of separated oscillatory field developed by Ramsey\cite{ramsey49}.
Oscillating magnetic fields of identical frequency are introduced at each
end of a homogeneous-field region. Spin-flip transitions are maximally induced
when the frequency of the oscillatory 
field is set at the resonance frequency corresponding to the
neutron precessing frequency. The neutron EDM
is determined from the shift of the resonance frequency when
the direction of the electric field is reversed.

Following the pioneering work of Purcell et al. at Oak Ridge in
1950\cite{smith51,smith57}, various improvements of the 
experimental techniques have been
introduced and similar experiments were carried out at
Oak Ridge\cite{miller67,dress68,baird69,dress73}, Brookhaven\cite{cohen69},
Bucharest\cite{apostolescu70}, Aldermaston, and Grenoble\cite{dress77}.
Table~\ref{tab:edmhis1} lists some
characteristics of these experiments. The 1977 measurement\cite{dress77}
at the Institut Laue-Langevin (ILL)
represented a four order-of-magnitude improvement in sensitivity
over the original Oak Ridge experiment. This was accomplished by
minimizing the statistical and systematic errors.
The statistical uncertainty in $d_n$ is:
\begin{equation}
\Delta d_n \propto \langle v \rangle / [E_0 L P (\phi_n t)^{1/2}].
\label{eq:larmor5}
\end{equation}
To obtain maximal sensitivity with a given running time $t$, 
the experiment needs to maximize
the electric field $E_0$, the distance $L$ between the RF coils, the
neutron polarization $P$, and the neutron flux $\phi_n$. In addition, the
mean neutron velocity $\langle v \rangle$ needs to be minimized.
Table~\ref{tab:edmhis1}
lists these parameters for various experiments.

Many sources of systematic errors have been identified. 
The $\vec v \times \vec E$ effect, also called the motional field effect,
refers to the additional magnetic field $\vec B_m$
viewed from the neutron rest frame,
\begin{equation}
\vec B_m = \frac {1}{c} \vec v \times \vec E_0,
\label{eq:larmor6}
\end{equation}
where $\vec v$ is the neutron velocity in the lab frame. If the electric
field $\vec E_0$ is not completely aligned with the magnetic field
$\vec B_0$, then $\vec B_m$ would acquire a non-zero component along
the direction of $\vec B_0$. For a cold neutron
of 100 m/sec, a misalignment angle of $1.5 \times 10^{-3}$ radians would lead
to an apparent neutron EDM of $10^{-23} e \cdot cm$.

As shown in Table~\ref{tab:edmhis1}, the most sensitive
neutron beam experiment\cite{dress77} obtained
$d_n < 3 \times 10^{-24} e \cdot cm$ with a systematic error of
$1.1 \times 10^{-24} e \cdot cm$.
The dominant contribution to the systematic error is the $\vec v \times \vec E$
effect, even though the misalignment angle is 
smaller than 
$1.1 \times 10^{-4}$ radians. The limitations from the $\vec v \times \vec E$
effect and from the magnetic field fluctuation can be removed
by using bottled UCN together with a comagnetometer, to be discussed next.

\subsection{Neutron EDM with ultra-cold neutrons}

There are two major limitations in the search for neutron EDM using 
thermal or cold neutron beams. First, the $\vec v \times \vec E$ 
systematic effect imposes
stringent requirements on the alignment of the $\vec E$ and $\vec B$
fields, as discussed earlier. Second, the transit time of neutron beams
in the magnetic spectrometer is relatively short, being $\sim 10^{-2}$
seconds. These limitations are responsible for the fact that
the best upper limit for neutron EDM achieved with the cold neutron beam at ILL
is $3 \times 10^{-24} e \cdot cm$, even though the statistical uncertainty
is at a lower level of $\sim 3 \times 10^{-25} e \cdot cm$.

In 1968 Shapiro first proposed\cite{shapiro68} using UCN in searches
for neutron EDM. The much lower velocities of UCNs will clearly
suppress the $\vec v \times \vec E$ effect. The amount of suppression
is further enhanced in an UCN bottle which allows randomization
of the neutron momentum directions. Another important advantage is that
the coherence time of UCN in a storage bottle will be of
the order $10^2 - 10^3$ seconds, a factor of $10^4 - 10^5$ improvement
over the neutron beam experiments. This significantly improves the
sensitivity for EDM signals relative to systematic effects.
An important price to pay, however, is the much lower flux for UCN relative
to that of cold neutron beams.
A series of neutron EDM experiments using UCN have been carried out at the
Petersburg Nuclear Physics Institute (PNPI) and at the ILL.

\subsubsection{Measurements at PNPI using UCN}

Immediately following Shapiro's proposal\cite{shapiro68}, preparation
for an UCN neutron EDM experiment started at PNPI. The early version
of the experiment~\cite{altarev80,altarev81}, used a 
``flow-through" type spectrometer with separated oscillating fields.
A 150 $cm^3$ liquid hydrogen moderator was used for UCN production,
and a constant magnetic field of 28 mG and an electric field 
of $\sim 25$ kV/cm were 
applied to the double-chamber of $\sim 20$ liters each. 
From four different sets of measurements, they obtained $d_n = (2.3 \pm 2.3)
\times 10^{-25} e \cdot cm$. At 90\% confidence level, $|d_n| < 6
\times 10^{-25} e \cdot cm$. 

Major modifications for the PNPI experiment were reported\cite{altarev86}
in 1986. 
Probably influenced by the ILL stored
UCN experiment\cite{pendlebury84}
reporting a confinement time of $\sim 60$ seconds, the PNPI group
modified their spectrometer to allow prolonged confinement of the
UCNs. They achieved a confinement time of $\sim 50$ seconds.
The result of this experiment was
$d_n = - (1.4 \pm 0.6) \times 10^{-25} e \cdot cm$, implying
$|d_n| < 2.6 \times 10^{-25} e \cdot cm$ at 95\% confidence level.

The most recent PNPI measurement was reported in 
1992\cite{altarev92,altarev96}. 
The result is
$d_n = [2.6 \pm 4.0 (stat.) \pm 1.6 (syst.)] \times 10^{-26} e \cdot cm$,
which corresponds to  $|d_n| < 1.1 \times 10^{-25} e \cdot cm$
at 95\% confidence level. Systematic errors appeared
to limit the sensitivity of this experiment to few times $10^{-26} e
\cdot cm$.

\subsubsection{Measurements at ILL using UCN}

Following the completion of the neutron EDM measurement\cite{dress77}
using the neutron beam magnetic resonance method, the interest
at ILL shifted to the use of UCN\cite{golub79}.
Unlike the PNPI group, the ILL group
started out with the UCN storage bottle technique and did not use the
flow-through technique. The first ILL result was
published in 1984\cite{pendlebury84}, which demonstrated the feasibility
of measuring neutron EDM with stored UCN. 

The sensitivity of the ILL measurement was significantly improved
in a subsequent experiment reported in 1990\cite{smith90}. A new
neutron turbine\cite{steyerl86} increased the UCN flux by a factor of 200
and a density of 10 UCN per $cm^3$ was achieved in the neutron
bottle. The electric field was raised to 16 kV/cm and the leakage
current was reduced from 50 nA to 5 nA. Following a three-year running
over 15 reactor cycles, the result was reported to be $d_n = - (3 \pm 5)
\times 10^{-26} e \cdot cm$, implying $|d_n| < 1.2 \times 10^{-25}
e \cdot cm$ at the 95 \% confidence level.

To overcome the systematic uncertainty caused by magnetic
field fluctuations in the UCN bottle, Ramsey suggested\cite{ramsey84}
the use of comagnetometers for EDM experiments. The idea was to store
polarized atoms simultaneously in the same bottle as the
neutrons. Fluctuation of the magnetic field will affect the spin
precession of the comagnetometer atoms, which can be monitored. The ILL
collaboration selected $^{199}$Hg as the comagnetometer. Effects
from the $^{199}$Hg EDM are negligible, since 
experiment\cite{romalis01}
showed that the EDM of $^{199}$Hg
was less than $2.1 \times 10^{-28} e \cdot cm$.

In 1991, an ILL experiment\cite{harris99} used a 20-liter UCN bottle
containing $3 \times 10^{10}/cm^3$ polarized $^{199}$Hg. The UCN coherence
time was 130 seconds, roughly a factor of two improvement over previous
experiment. However, the maximum electric field in this UCN
bottle is only 4.5 kV/cm, roughly a factor of 3.5 lower than before.
The UCN flux also appeared to be a factor of 4 lower than in the earlier
experiment. Data were collected over ten reactor cycles of 50 days' length,
and the $^{199}$Hg comagnetometer was shown to reduce effects from
magnetic field fluctuations significantly. The result of this
experiment was $d_n = (1.9 \pm 5.4) \times 10^{-26} e \cdot cm$.
An upper limit on the neutron EDM of $|d_n| < 9.4 \times 10^{-26} e \cdot cm$
was obtained at 90\% confidence level. When this result was combined with
the result from the earlier ILL experiment\cite{smith90}, an improved
upper limit of $6.3 \times 10^{-26} e \cdot cm$ was obtained. 

The most recent ILL experiment\cite{baker06} reached a higher $E$ field
of 10 KV/cm by adding $10^{-3}$ torr of helium gas to prevent spark.
An important systematic effect due to geometric phase, which arises
when the trapped particles experience a magnetic field gradient $\partial
B_z/\partial z$, was identified\cite{pendlebury04} and corrected for.
The upper limit of neutron EDM is now pushed down to 
$2.9 \times 10^{-26} e \cdot cm$ at 90\% confidence level.

The ILL experiments demonstrated the advantage of using a comagnetometer
for reducing a dominant source of systematic error.
It is conceivable that the sensitivity on neutron EDM can be further improved
if more intense UCN flux together with a suitable comagnetometer become 
available. New experiments have been proposed at
ILL and the SNS using UCN produced in superfluid $^4$He, as discussed in the
next Section.

\section{Future Neutron EDM Experiments}

Several new neutron EDM experiments aiming at improved sensitivities have been
proposed\cite{raidal08}. To achieve a greater statistical 
accuracy, it is important to
increase the UCN flux, the electric field strength, and the UCN storage
time. Golub and Pendlebury\cite{golub75} 
first suggested that higher UCN flux/density can
be obtained using the down-scattering processes, where a fraction of an
intense cold neutron beam scatter inelastically from a suitable
material and lose practically all their energies to become UCNs. An 
intense UCN source based on down-scattering process in solid deuterium
has been constructed at Los Alamos\cite{saunders04}, 
and another one is being constructed\cite{psi}
at the Paul-Scherrer Institute (PSI). 

A different technique\cite{golub77} using superfluid
helium has also been utilized to produce intense UCNs 
stored in bottles\cite{huffman00}.
The energy-momentum dispersion curves for He-II and free neutrons intersect
at the neutron momentum of $\sim$ 8.9 \AA. Therefore, monochromatic
neutron beam at this momentum can efficiently produce UCNs by exciting
phonons in He-II. An EDM experiment using UCN produced in superfluid
helium has several important advantages. First, a high electric field 
can be applied due to the high dielectric constant of liquid helium.
Second, more uniform magnetic field can be obtained with superconducting
shields. Finally, the storage time can be much improved since the loss
from wall-scattering is greatly reduced due to the low temperature of the
walls.

In order to benefit from the projected improvement in statistical accuracy
for future neutron EDM experiments, systematic uncertainties have to be 
reduced accordingly. Most of the systematic 
uncertainties are related to the stability
and uniformity of the magnetic fields. As discussed later, various schemes
of active and passive magnetometers, as well as co-magnetometers have been
proposed for future neutron EDM experiments. We now describe the main 
features and status of these future neutron EDM experiments.

\subsection{Room Temperature Experiments at PSI and ILL}

A new experiment\cite{psiedm}
based on the Sussex-RAL-ILL apparatus has been proposed at PSI. This new
experiment plans to use the high intensity UCN source currently being 
constructed at PSI\cite{psi}. This UCN source 
is expected to deliver UCN densities
of $\sim 1000$ per cm$^3$ to the EDM experiment, roughly two orders
of magnitude better than the existing ILL experiment. 

Extensive R\&D efforts are underway to study various schemes for improving
the appararus. An array of laser-pumped Cs 
magnetometers\cite{groeger05} placed near the UCN
cell will monitor the magnetic field and its gradients. These magnetometers
could provide inputs for the correction coils to actively stabilize the 
magnetic field and minimize the field gradients. The sensitivity of the
$^{199}$Hg co-magnetometer could also be improved, possibly from a sensitivity
of 200 femto-Tesla to 40 femto-Tesla\cite{kirch06}. 
Addition of a second co-magnetometer
such as $^3$He and $^{129}$Xe has also been considered. Since different
co-magnetometers have different sensitivities to the geometric-phase effect,
the additional co-magnetometer could 
help to isolate this effect\cite{raidal08}. 

Using the improved version of the Sussex-RAL-ILL apparatus together with the
intense PSI UCN source, it is anticipated that a sensitivity of $5 \times
10^{-27}$ $e \cdot cm$ can be reached after data-taking during 2009-2010. 
Meanwhile, a design of a new apparatus is underway, which can lead to a 
factor of 10 improvement in sensitivity due to a larger experimental volume, 
an improvement in the electric field strength, a better match to the UCN 
source, and a longer running time. Data-taking for this new apparatus is
planned for 2011-2014\cite{raidal08}.

Another room-temperature UCN experiment, led by the PNPI group, will
use an apparatus consisting of 4 back-to-back measurement chambers with
opposite electric fields. This design allows cancellation of some
systematic errors. A total of 16 Cs magnetometers will be used for
magnetic field stabilization. The experiment plans to use the UCN source 
at ILL to reach a sensitivity of $10^{-27}$ $e \cdot cm$.
 
\subsection{ILL Cryogenic Experiment}

A CryoEDM experiment by the Sussex/RAL/Oxford/Kure/ILL collaboration
is being installed at ILL\cite{balashov07}. 
UCNs will be produced in superfluid $^4$He 
cell with cold neutron beam. Production of UCN with this technique 
has been demonstrated\cite{baker03} at ILL. A 400 KV high 
voltage supply will be connected
to the HV electrode. The holding
field $B_0$ will increase by a factor of 5 over the latest ILL
experiment to reduce the geometric-phase effect, which is proprotional 
to $1/B_0^2$. Neutrons will be detected using $^6$LiF-coated silicon 
solid-state detectors\cite{baker03a} placed
inside the 0.5 K He-II liquid via the n + $^6$Li $\to$ $^3$H + $^4$He 
reaction.

Since $^{199}$Hg co-magnetometer can not 
function at low temperature, alternative
techniques are required to monitor the magnetic field.
An axial shielding factor of $\sim 10^6$ will be obtained
from mu-metal, superconducting, and active shieldings. An array of
12 pickup loops for a SQUIDS system are placed behind the grounded 
electrodes to monitor the magnetic fields. A control cell adjacent to
the measurement cell will have no applied electric field and act as
a neutron magnetometer.

This experiment expects to obtain a statistical sensitivity of $\sim 
10^{-27} e \cdot cm$ around end of 2008. With further improvement
of the apparatus and a new beam line at ILL with six times higher
intensity of 8.9 A neutrons, a sensitivity of $\sim 2 \times 10^{-28}
e \cdot cm$ is anticipated after two to three years of running.

\subsection{SNS nEDM experiment}

\begin{figure}[th]
\centerline{\psfig{file=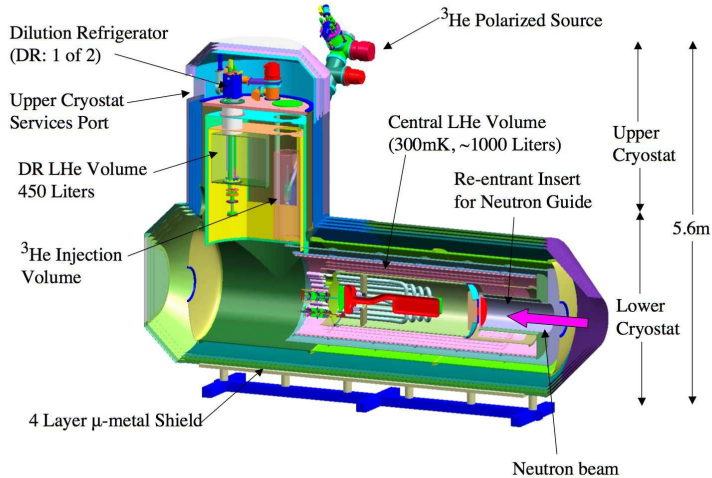,width=4.3in}}
\caption{Schematics of the SNS nEDM apparatus.}
\label{fig:sns}
\end{figure}

The nEDM experiment\cite{nedm,ito07}, based on the idea 
outlined by Golub and Lamoreaux\cite{golub94} in
1994, will run at the 8.9 \AA~Fundamental Neutron Physics Beamline (FNPB) 
at the Spallation Neutron Source (SNS) at the Oak Ridge National Laboratory.
The schematics of the proposed apparatus is shown in Figure~\ref{fig:sns}.
Like the ILL CryoEDM experiment, superfluid $^4$He will be used to produce
and trap the UCNs. However, a major difference for the nEDM experiment
is the use of polarized $^3$He as a co-magnetometer as well as a spin
analyser.

In the proposed neutron EDM experiment, a small concentration of
polarized ${}^{3}$He atoms ($X \sim 10^{-10}$) would be introduced
into the superfluid to serve as a comagnetometer. The ${}^{3}$He atoms
would also function as a highly sensitive spin analyzer due to the
large difference between the $n$-${}^{3}$He absorption with total spin
$J=0$ compared to $J=1$.  The absorption reaction $n\
+ {}^{3}\text{He}\to p\ +{}^{3}\text{H}\ $ releases 764~keV of total
kinetic energy.  This recoil energy excites short-lived molecules in
the superfluid ${}^{4}$He which emit ultraviolet scintillation
light.  Consequently, the observed rate of
scintillations depends on the relative angle between the UCN and
${}^{3}$He spins.  In a transverse magnetic field $B_0$, the UCN and
${}^{3}$He spins will precess at their respective Larmor frequencies:
$\om_n = \gam_n B_0$, and $\om_3 = \gam_3 B_0$ where $\gam_i$ is the
gyromagnetic ratio of each species. If the ${}^{3}$He and UCN spins
are parallel at time $t=0$, a relative angle between the spins
develops over time because the ${}^{3}$He magnetic moment is larger
than that of the neutron ($\gam_3 \approx 1.1\, \gam_n$).
In the presence of a static electric field $E$ parallel to $B_0$,
the rate of scintillations observed is modulated at the difference of
the two spin precession frequencies:
\begin{equation} \om_\text{rel} = (\gam_3 - \gam_n)B_0
+\, 2 d_n E/\hbar.
\label{eq:omegarelwithE}
\end{equation}

\vfill
\eject

Eq.~\ref{eq:omegarelwithE} shows that $\om_\text{rel}$ depends only on
$d_n E$ in the limit of $B_0 \to 0$. Alternatively, the experimental
signal would become independent of $B_0$ if the condition $\gam_3 -
\gam_n = 0$ were satisfied. Spurious signals due to inhomogeneity or
slow drifts in the magnetic fields would thereby be eliminated. The
UCN and $^{3}$He magnetic moments can be modified, and in fact
equalized, by the dressed spin effect\cite{golub94,haroche70}
in which a particle's effective magnetic moment is
modified by applying an oscillating magnetic field $B_d \cos \om_d t$
perpendicular to $B_0$. In the weak-field limit ($B_0 \ll \om_d/\gam$),
the dressed magnetic
moment $\gam_i'$ is given by
\begin{equation}\gam_i' = \gam_i J_0(x_i), \qquad x_i \equiv \gamma_i
B_d/\om_d,
\label{eq:J0relation}
\end{equation}
where $J_0$ is the zeroth-order Bessel function. Using this expression, 
one can solve for the ``critical''
dressing field magnitude which makes $\gam_n' = \gam_3'$. If this
critical dressing field is applied, corresponding to $x_3 =
1.32$, the relative precession between the UCN and
${}^{3}$He (Eq.~\ref{eq:omegarelwithE}) vanishes except for the
contribution from $d_n E$. 
 
Extensive R\&D effort has been underway. In particular, the distribution
of $^3$He in superfluid $^4$He and the $^3$He diffusion coefficient at
temeprature below 1 K have been 
measured\cite{lamoreaux02,hayden03,hayden04}. The dielectric strength of
superfluid helium has been studied using a prototype 
test apparatus\cite{long06}.
The relaxation of polarized $^3$He in a mixture of $^3$He and $^4$He
at a temperature below the $\lambda$ point 
has been investigated\cite{ye06}.
A study of the dressed-spin effect has been carried out using a
cold atomic $^3$He source of 99.5\% polarization constructed for
the nEDM experiment\cite{esler07}. 
Optimization of the neutron beam line has
also been carried out\cite{ito06}. 

The FNPB is currently under construction and is scheduled to be completed
in 2010. The construction of the nEDM experiment is expecetd to begin
in 2009. The goal is a two order-of-magnitude improvement
in sensitivity over the present limit.

\end{document}